\documentclass[epj]{svjour}
\usepackage{textcomp}
\usepackage{amssymb}
\usepackage[fleqn]{amsmath}
\usepackage{graphics}
\begin{document}
\title{A single-oscillator quantum model for magnetochiral birefringence}
\author{M. Donaire \inst{1,2} \and G.L.J.A. Rikken \inst{2}  \and B.A. van Tiggelen \inst{1}
}                     
\offprints{}          
\institute{ Universit\'{e} Grenoble 1/CNRS, LPMMC UMR 5493, B.P.
166, 38042 Grenoble, France  \and LNCMI, UPR 3228 CNRS/INSA/UJF Grenoble 1/UPS, Toulouse
\& Grenoble, France}
\date{Received: date / Revised version: date}

\abstract{We derive an analytical expression for the magnetochiral birefringence of a dilute diamagnetic chiral molecular medium subjet to a constant magnetic field.
We use the single-oscillator model of Condon \emph{et al.} \cite{Condon1,Condon2} to describe the optical properties of the individual molecules.
The result is a function of the refractive index and the rotatory power. This result is compared to experimental data.}

\maketitle


\section{Introduction}
\label{Intro}
Magnetochiral effects were first predicted fifty years ago \cite{Groenewege,Portigal}. They result from the joined effect of the natural optical 
activity (NOA) and the magneto-optical activity (MOA) of chiral molecules under the action of a static magnetic field. Physically, they are manifestations of the
simultaneous breaking of fundamental
symmetries of nature, namely spatial inversion and time-reversal. Unlike NOA and Faraday rotation, the magnetochiral effects are independent 
of the polarization state of light and depend only on the relative orientation between the light wave vector, $\mathbf{k}$, and the static magnetic field, $\mathbf{B_{0}}$.\\ 
\indent In emission, the fundamentals of magnetochiral dichroism have been explained  in Ref.\cite{Barron84,Wagniere82}. It is characterized
by the variation of the absorption coefficient of a given enantiometer spiece when the orientation of the probe light is reversed with 
respect to $\mathbf{B_{0}}$.\\
\indent In refraction, the theoretical basis of magnetochiral birefringence in molecules has been discussed first in Ref.\cite{Barron84,Baranova77}. It is quantified 
by the magnetochiral index, $\Delta n_{MCh}$, which is the difference between the refractive index 
for unpolarized light propagating parallel to $\mathbf{B_{0}}$, $n_{\upuparrows}$, and antiparallel to $\mathbf{B_{0}}$, $n_{\uparrow\downarrow}$, 
$\Delta n_{MCh}= n_{\upuparrows}-n_{\uparrow\downarrow}$.\\
\indent Magnetochiral dichroism was confirmed experimentally in 1997 \cite{Rikken97}, while birefringence was experimentally observed a year 
later \cite{Kleindienst98}. In this article we concentrate on birefringence and calculate the magnetochiral anisotropy in the transparent region of the spectrum.\\ 
\indent There are three basic approaches to this problem. The first one, due to Baranova and Zel'dovich \cite{Baranova79}, is based on the Becquerel's formula for the Faraday
 effect. It yields $\Delta n_{MCh}$ as a function of the NOA coefficient. More recently, \emph{ab initio} calculations based on Hartree-Fock methods and using quantum models of optical activity \cite{Wagniere82,Barron84} have been 
developed \cite{Coriani2002,Jansik2006}. Also, classical dipole-dipole interaction models  which use empirical polarizability data have been 
succesful to some extent \cite{Applequist73,Pinheiro,Ruchon2006}.\\
\indent Here we follow an alternative approach. We extend the single-oscillator quantum model of Condon, Altar and Eyring \cite{Condon1,Condon2} developed 
to describe molecular rotatory power, to account for magnetochiral anisotropy upon including orbital magnetism by means of a Zeeman potential. 
In Ref.\cite{Condon2} the authors 
computed the rotatory power of a chiral compound using the microscopical data of the spatial distribution of charges and energy levels within the molecular units. 
They found their results to be extremely sensitive to the details in these data, which implies large uncertainties if not all details are known. In order to avoid these difficulties and still keep the 
essenctials of the quantum model, our approach is not microscopical but rather semi-empirical. On the one hand we stick to the quantum model to derive a relation 
between magnetochiral anisotropy and NOA. On the other hand, we fit the effective parameters of the model using the experimental values of the refractive index 
and the rotatory power instead of microscopical data.

\section{The model}
\label{Model}
We adapt the single-oscillator quantum model of Condon, Altar and Eyring \cite{Condon1,Condon2} to which we add an external static magnetic field 
$\mathbf{B}_{0}$. According to that model the optical activity of a single molecule is determined by a chromophoric electron within a 
 chiral center. We further simplify the chiral center using a two-particle system in which the chromophoric electron of 
charge $q_{e}=-e$ and mass $m_{e}$ is bound to a nucleus of effective charge $q_{N}=e$ and mass $m_{N}\gg m_{e}$. The binding interaction is modeled by a
harmonic oscillator potential, $V^{HO}=\frac{\mu}{2}(\omega_{x}^{2}x^{2}+\omega_{y}^{2}y^{2}+\omega_{z}^{2}z^{2})$, 
to which a term $V_{C}=C\:xyz$ is added to break the mirror symmetry at leading order in perturbation theory. The coordinates $x$, $y$, $z$ are those of the relative position vector, 
$\mathbf{r}=\mathbf{r}_{N}-\mathbf{r}_{e}$, and $\mu=m_{N}m_{e}/M$ with $M=m_{N}+m_{e}$. The center of mass position vector is 
$\mathbf{R}=(m_{N}\mathbf{r}_{N}+m_{e}\mathbf{r}_{e})/M$. The principal axis of the oscillator as well as the pseudoscalar 
parameter $C$ can be derived from the Coulomb interaction of the chiral center with the (chiral) distribution of charges within the rest of atomic groups of the 
molecule. In our case we will fit the unknown parameters \emph{a posteriori} using macroscopic experimental data.\\
\indent When an external uniform and constant magnetic field
$\mathbf{B}_{0}$ is applied  and the molecule is exposed to a classical probe field, the total Hamiltonian of the system reads, $H=H_{0}+H^{0}_{EM}+W$, 
with $H^{0}_{EM}$ the electromagnetic (EM) energy of the free optical field and 
\begin{eqnarray}
H_{0}&=&\sum_{i=e,N}\frac{1}{2m_{i}}[\mathbf{p}_{i}-q_{i}\mathbf{A}_{0}(\mathbf{r}_{i})]^{2}+V^{HO}+V_{C},\label{H0}\\
W&=&\sum_{i=e,N}\frac{-q_{i}}{m_{i}}[\mathbf{p}_{i}-q_{i}\mathbf{A}_{0}(\mathbf{r}_{i})]\cdot\mathbf{A}(\mathbf{r}_{i},t)
 +\frac{q^{2}_{i}}{2m_{i}}\textrm{A}^{2}(\mathbf{r}_{i}).\label{wed}
\end{eqnarray}
In the EM vector potential we have separated the contribution of the external static field, $\mathbf{A}_{0}(\mathbf{r}_{i})=\frac{1}{2}\mathbf{B}_{0}\wedge\mathbf{r}_{i}$,
from the one of the optical probe field, $\mathbf{A}(\mathbf{r}_{i},t)$. For simplificity we take $\mathbf{A}(\mathbf{r}_{i})$ to to be real and monochromatic, 
$\mathbf{A}(\mathbf{r}_{i},t)=\frac{1}{2}\mathbf{A}_{\omega}[e^{i\omega t-i\mathbf{k}\cdot\mathbf{r}_{i}}+e^{-i\omega t+i\mathbf{k}\cdot\mathbf{r}_{i}}]$, with
$\mathbf{A}_{\omega}$ constant and real.\\ 
\indent Considering the molecule at rest, the unitary operator 
U$=\exp{[-i\frac{e}{2\hbar}(\mathbf{B}_{0}\wedge\mathbf{r})\cdot\mathbf{R}]}$ maps the Hamiltonian $H_{0}$ into
$\tilde{H}_{0}=$U$^{\dagger}H_{0}$U, which depends only on the internal motion \cite{Herold,Dippel},
\begin{equation}
\tilde{H}_{0}=\frac{1}{2\mu}\mathbf{p}^{2}+V^{HO}+V_{C}+V_{Z}+\mathcal{O}(\textrm{B}_{0}^{2}).\label{effective}
\end{equation}
In the above equation $\mathbf{p}$ is the conjugate momentum of $\mathbf{r}$, 
$\mathbf{p}=\mu(\mathbf{p}_{N}/m_{N}-\mathbf{p}_{e}/m_{e})$, and $V_{Z}=\frac{e}{2\mu^{*}}(\mathbf{r}\wedge\mathbf{p})\cdot\mathbf{B}_{0}$ 
is the Zeeman potential with $\mu^{*}=\frac{m_{N}m_{e}}{m_{N}-m_{e}}$. $V_{Z}$ guarantees the break down of time-reversal at first order perturbation theory.
The ground state of the Hamiltonian $\tilde{H}_{0}$, perturbed by both magnetic field and chirality up to order $C$B$_{0}$, reads
\begin{align*}
|\tilde{\Omega}_{0}\rangle&=|0\rangle-\mathcal{C}|111\rangle
-i\mathcal{B}_{0}^{z}\eta^{yx}|110\rangle\nonumber\\
&+i\mathcal{B}_{0}^{z}\mathcal{C}\eta^{yx}\left(|001\rangle+2|221\rangle\right)\nonumber\\
&-\sqrt{2}i\mathcal{B}_{0}^{z}\mathcal{C}\left(\frac{2\omega_{x}-\omega_{z}\eta^{yx}}{\omega_{z}+2\omega_{x}}|201\rangle
-\frac{2\omega_{y}+\omega_{z}\eta^{yx}}{\omega_{z}+2\omega_{y}}|021\rangle\right)\nonumber\\
&+\sum\textrm{ cyclic permutations}.\nonumber
\end{align*}
Correspondingly, the ground state of $H_{0}$ is $|\Omega_{0}\rangle=$U$|\tilde{\Omega}_{0}\rangle$ and  
for simplicity we consider the center of mass fixed at $\langle\mathbf{R}\rangle=\mathbf{0}$.
In the above equation the states $|n_{x} n_{y} n_{z}\rangle$ refer to the eigenstates of the harmonic oscillator Hamiltonian. The dimensionless parameters
are, 
\begin{eqnarray}
\mathcal{C}&=&\frac{C\hbar^{1/2}}{(2\mu)^{3/2}(\omega_{x}+\omega_{y}+\omega_{z})(\omega_{x}\omega_{y}\omega_{z})^{1/2}},\nonumber\\ 
\mathcal{B}_{0}^{z}&=&\frac{eB_{0}^{z}}{4\mu^{*}\sqrt{\omega_{x}\omega_{y}}},\qquad \eta^{yx}=\frac{\omega_{y}-\omega_{x}}{\omega_{y}+\omega_{x}},
\end{eqnarray}
and analogous definitions hold for the cyclic permutations of the last two parameters. 
The $\eta$ factors are assumed to be small quantities which quantify the anisotropy of the oscillator. 
In the following, all our calculations restrict to the lowest order in $\mathcal{C}$, $\mathcal{B}_{0}^{i}$ and $\eta^{ij}$. It will be shown that the dimensionless 
combination $\mathcal{C}\eta^{zy}\eta^{yx}\eta^{xz}$ determines the chiral observables. This quantity and the average frequency, 
$\omega_{0}=(\omega_{x}+\omega_{y}+\omega_{z})/3$, are the only unknown parameters to be determined from experimental data. 

\section{Constitutive equations}
\label{Constitution}
In order to compute $\Delta n_{MCh}$ we calculate first the linear response of the system to the optical field. To this end we apply first order time-dependent perturbation 
theory to the ground state $\Omega_{0}$ and compute the expectation value of the electric dipole moment, $\langle \mathbf{d}\rangle$, magnetic dipole moment, $\langle \mathbf{m}\rangle$, 
and electric quadrupole moment operators, $\langle\bar{\mathbf{\Theta}}\rangle$. As functions of $\mathbf{r}$, $\mathbf{p}$, the 
corresponding quantum operators read,
\begin{equation*}
\mathbf{d}=e\mathbf{r},\quad\overline{\mathbf{\Theta}}=\sum_{i=e,N}q_{i}\mathbf{r}_{i}\otimes\mathbf{r}_{i}=-\frac{e\mu}{\mu^{*}}\mathbf{r}\otimes\mathbf{r}+e(\mathbf{r}\otimes\mathbf{R}+\mathbf{R}\otimes\mathbf{r}),\nonumber\\
\end{equation*}
\begin{equation*}
\mathbf{m}=\sum_{i=e,N}\frac{q_{i}}{2}\mathbf{r}_{i}\wedge\mathbf{p}_{i}=\frac{e}{2}(-\mu^{*-1}\vec{\mathbf{r}}\wedge\vec{\mathbf{p}}+M^{-1}\vec{\mathbf{r}}\wedge \vec{\mathbf{P}}+\mu^{-1}\vec{\mathbf{R}}\wedge\vec{\mathbf{p}})\nonumber.
\end{equation*}
The expectation value of the $\mathbf{R}$-dependent terms vanish for $\langle\mathbf{R}\rangle=\mathbf{0}$. 
For the calulation of the remaining terms we use the U-transformed states and operators. The operators $\mathbf{d}$ and $\bar{\mathbf{\Theta}}$ are U-invariant, while the magnetic dipole transforms as 
$\tilde{\mathbf{m}}=\frac{-e}{2\mu^{*}}\mathbf{r}\wedge\mathbf{p}-\frac{e^{2}}{4M}\vec{\mathbf{r}}\wedge(\mathbf{B}_{0}\wedge\mathbf{r})$
plus terms of null contribution for $\langle\mathbf{R}\rangle=\mathbf{0}$. 
Using the commutation relations of the Hamiltonian $H_{0}$ with $\mathbf{r}$ and $\mathbf{R}$ to replace the momentum operators in Eq.(\ref{wed}), U-transforming
and expanding the exponentials up to $\mathcal{O}(\mathbf{k}\cdot\mathbf{r})$ and separating positive and negative frequency modes, we obtain 
for the time-evolution of the perturbing optical interaction (\ref{wed}),
\begin{eqnarray}
\tilde{W}(t)&=&\frac{1}{2}[\tilde{W}^{+}e^{i\omega t}+\tilde{W}^{-}e^{-i\omega t}],\quad \textrm{with }\nonumber\\
\tilde{W}^{+}&\simeq&-\frac{ie}{\hbar}\mathbf{A}_{\omega}\cdot[\tilde{H}_{0},\mathbf{r}]+\frac{e\mu}{\hbar\mu^{*}}\mathbf{A}_{\omega}\cdot[\tilde{H}_{0},\mathbf{r}]
(\mathbf{k}\cdot\mathbf{r}),\nonumber\\
\tilde{W}^{-}&\simeq&-\frac{ie}{\hbar}\mathbf{A}_{\omega}\cdot[\tilde{H}_{0},\mathbf{r}]-\frac{e\mu}{\hbar\mu^{*}}\mathbf{A}_{\omega}\cdot[\tilde{H}_{0},\mathbf{r}]
(\mathbf{k}\cdot\mathbf{r}).\nonumber
\end{eqnarray}
 For any multipole moment operator $\mathbf{O}$, its time dependent 
expectation value in the ground state of the system reads,
\begin{eqnarray}
\langle\mathbf{O}\rangle(t)&=&\Re\sum_{j}
\langle\tilde{\Omega}_{0}|\tilde{\mathbf{O}}|\Phi_{j}
\rangle\Bigl[\frac{\langle\Phi_{j}|\tilde{W}^{+}|\tilde{\Omega}_{0}\rangle e^{i\omega t}}{E_{j}-3\hbar\omega_{0}/2+\hbar\omega}\nonumber\\
&+&\frac{\langle\Phi_{j}|\tilde{W}^{-}|\tilde{\Omega}_{0}\rangle e^{-i\omega t}}{E_{j}-3\hbar\omega_{0}/2-\hbar\omega}
\Bigr],\label{lao}
\end{eqnarray}
where the intermediate states $\{|\Phi_{j}\rangle\}$ are eigenstates of $\tilde{H}_{0}$ up to order B$_{0}C$. We use the closure relation
\begin{equation*}
\sum_{j}|\Phi_{j}\rangle \frac{1}{E_{j}-3\omega_{0}/2\pm\hbar\omega}\langle\Phi_{j}|=\frac{1}{H^{HO}+V_{C}+V_{Z}-3\omega_{0}/2\pm\hbar\omega}\nonumber
\end{equation*}
in Eq.(\ref{lao}), where we must expand the r.h.s. of the above relation around $(H^{HO}-3\omega_{0}/2\pm\hbar\omega)^{-1}$ up to order $V_{C}V_{Z}$. The computation of 
$\langle\mathbf{O}\rangle(t)$ is actually of up to third-order in perturbation theory, with the total perturbative potential $(V_{C}+V_{Z}+\tilde{W})$ acting 
on $|0\rangle$ and retaining terms up to order one in $V_{C}$, $V_{Z}$ and $\tilde{W}$. The expectation value of the dipole and quadrupole moments read,
\begin{eqnarray}
\langle\mathbf{d}\rangle&=&\Re\sum_{j}\langle
\tilde{\Omega}_{0}|e\mathbf{r}|\Phi_{j}\rangle\frac{e^{i\omega t}}{\hbar\omega+E_{j}-3\omega_{0}/2}\langle\Phi_{j}|\nonumber\\
&\times&\frac{-ie}{\hbar}[\tilde{H}_{0},\mathbf{r}]\cdot\mathbf{A}_{\omega}|\tilde{\Omega}_{0}\rangle+[\omega\rightarrow-\omega]\label{e1}\\
&+&\Re\sum_{j}\langle
\tilde{\Omega}_{0}|e\mathbf{r}|\Phi_{j}\rangle\frac{e^{i\omega t}}{\hbar\omega+E_{j}-3\omega_{0}/2}\langle\Phi_{j}|\nonumber\\
&\times&\frac{e\mu}{2\hbar\mu^{*}}\mathbf{r}\wedge[\tilde{H}_{0},\mathbf{r}]|\tilde{\Omega}_{0}\rangle
\cdot(\mathbf{k}\wedge\mathbf{A}_{\omega})-[\omega\rightarrow-\omega]\nonumber\\
&+&\Re\sum_{j}\langle
\tilde{\Omega}_{0}|e\mathbf{r}|\Phi_{j}\rangle\frac{e^{i\omega t}}{\hbar\omega+E_{j}-3\omega_{0}/2}\langle\Phi_{j}|\nonumber\\
&\times&\frac{e\mu}{2\hbar\mu^{*}}\mathbf{k}\cdot[\tilde{H}_{0},\mathbf{r}\otimes\mathbf{r}]\cdot\mathbf{A}_{\omega}\tilde{\Omega}_{0}\rangle-[\omega\rightarrow-\omega],\nonumber\\
\langle\mathbf{m}\rangle&=&\Re\sum_{j}\langle
\tilde{\Omega}_{0}|\tilde{\mathbf{m}}|\Phi_{j}\rangle\frac{e^{i\omega t}}{\hbar\omega+E_{j}-3\omega_{0}/2}\langle\Phi_{j}|\nonumber\\
&\times&\frac{e\mu}{2\hbar\mu^{*}}\mathbf{r}\wedge[\tilde{H}_{0},\mathbf{r}]\cdot(\mathbf{k}\wedge\mathbf{A}_{\omega})|\tilde{\Omega}_{0}\rangle-[\omega\rightarrow-\omega]\nonumber\\
&+&\Re\sum_{j}\langle
\tilde{\Omega}_{0}|\tilde{\mathbf{m}}|\Phi_{j}\rangle\bigl[\frac{e^{i\omega t}}{\hbar\omega+E_{j}-3\omega_{0}/2}\langle\Phi_{j}|\label{e2}\\
&\times&\frac{-ie}{\hbar}[\tilde{H}_{0},\mathbf{r}]\cdot\mathbf{A}_{\omega}|\tilde{\Omega}_{0}\rangle+[\omega\rightarrow-\omega],\nonumber\\
\langle\overline{\mathbf{\Theta}}\rangle&=&\Re\sum_{j}\langle
\tilde{\Omega}_{0}|\frac{-e\mu}{\mu^{*}}\mathbf{r}\otimes\mathbf{r}|\Phi_{j}\rangle\frac{e^{i\omega t}}{\hbar\omega+E_{j}-3\omega_{0}/2}\langle\Phi_{j}|\nonumber\\
&\times&\frac{-ie}{\hbar}[\tilde{H}_{0},\mathbf{r}]\cdot\mathbf{A}_{\omega}|\tilde{\Omega}_{0}\rangle+[\omega\rightarrow-\omega].\label{e3}
\end{eqnarray}
We adapt and extend appropriately the nomenclature of Ref.\cite{Barron84} in favour of SI units to write these relations in the following form,
\begin{eqnarray}
\langle\textrm{d}^{\omega}_{i}\rangle&=&\alpha_{ij}\textrm{E}^{j}_{\omega} +\alpha^{'}_{ijk}\textrm{B}^{j}_{0}\dot{\textrm{E}}^{k}_{\omega} 
-G_{ij}\dot{\textrm{B}}^{j}_{\omega} +G^{'}_{ijk}\textrm{B}^{j}_{0}\textrm{B}^{k}_{\omega} \nonumber\\
&+&\frac{1}{2}A_{ijk}\textrm{k}^{j}\textrm{E}^{k}_{\omega}+\frac{1}{2}A^{'}_{ijkl}\textrm{B}^{l}_{0}\textrm{k}^{j}\textrm{E}^{k}_{\omega} ,\nonumber\\
\langle\textrm{m}^{\omega}_{i}\rangle&=&M_{ij}\textrm{B}^{j}_{\omega} +M^{'}_{ijk}\textrm{B}^{j}_{0}\dot{\textrm{B}}^{k}_{\omega} 
+G_{ij}\dot{\textrm{E}}^{j}_{\omega} -G{'}_{ijk}\textrm{B}^{j}_{0}\textrm{E}^{k}_{\omega} ,\nonumber\\
\langle\Theta^{\omega}_{ij}\rangle&=&iA_{kij}\textrm{E}^{k}_{\omega}+iA^{'}_{kijl}\textrm{B}^{l}_{0}\textrm{E}^{k}_{\omega} ,
\end{eqnarray}
where $\mathbf{E}_{\omega}$ and $\mathbf{B}_{\omega}$ are the complex electric and magnetic fields at the center of mass, 
$\mathbf{E}_{\omega}=i\omega\mathbf{A}_{\omega}e^{-i\omega t}$, $\mathbf{B}_{\omega}=i\mathbf{k}\wedge\mathbf{A}_{\omega}e^{-i\omega t}$, and 
the dipole and quadrupole moments are complex-valued as well. 
The rotational average of the above equations --denoted by the subscript rot-- is carried out following the usual prescription \cite{Craigbook}. 
2-rank tensors average to tensors proportional to Kronecker's delta, 3-rank tensors average to 
tensors proportional to Levi-Civita's tensor and 4-rank tensors average to a sum of tensor products of two Kronecker's deltas. This gives,
\begin{eqnarray}
\langle\textrm{d}^{\omega}_{i}\rangle_{\textrm{rot}}&=&\alpha_{E}\delta_{ij}\textrm{E}^{j}_{\omega} +\chi\epsilon_{ijk}\textrm{B}^{j}_{0}\dot{\textrm{E}}^{k}_{\omega} 
-\beta\delta_{ij}\dot{\textrm{B}}^{j}_{\omega} +\gamma\epsilon_{ijk}\textrm{B}^{j}_{0}\textrm{B}^{k}_{\omega} \nonumber\\
&+&\frac{1}{2}\xi[(\mathbf{B}_{0}\cdot\mathbf{k})\textrm{E}_{i}^{\omega}+(\mathbf{B}_{0}\cdot\mathbf{E}_{\omega})\textrm{k}_{i}],\nonumber\\
\langle\textrm{m}^{\omega}_{i}\rangle_{\textrm{rot}}&=&\alpha_{M}\delta_{ij}\textrm{B}^{j}_{\omega} +\zeta\epsilon_{ijk}\textrm{B}^{j}_{0}\dot{\textrm{B}}^{k}_{\omega} 
+\beta\delta_{ij}\dot{\textrm{E}}^{j}_{\omega} -\gamma\epsilon_{ijk}\textrm{B}^{j}_{0}\textrm{E}^{k}_{\omega} ,\nonumber\\
\langle\Theta^{\omega}_{ij}\rangle_{\textrm{rot}}&=&i\xi(\textrm{E}_{i}^{\omega}\textrm{B}^{0}_{j}
+\textrm{E}_{j}^{\omega} \textrm{B}^{0}_{i})+i\varsigma(\mathbf{B}_{0}\cdot\mathbf{E}_{\omega})\delta_{ij},
\end{eqnarray}
where $\alpha_{E}$  and $\alpha_{M}$ are the electric and magnetic polarizabilities respectively. 
$\chi$ and $\zeta$ describe the electric and magnetic Faraday 
effects respectively, $\beta$ is the molecular rotatory factor responsible for 
the natural optical activity and $\gamma$ and the $\xi$ factors give rise to the magnetochiral 
anisotropy. Note that the dipole-quadrupole polarizability does not survive rotational average unless $\mathbf{B}_{0}\neq\mathbf{0}$.

\section{Results}
\label{Results}
At leading order in the anisotropy factors, the polarizabilities that follow from Eqs.(\ref{e1}-\ref{e3}) are,
\begin{eqnarray}
\alpha_{E}&=&\frac{e^{2}}{\mu(\omega_{0}^{2}-\omega^{2})},\quad\alpha_{M}=\frac{4e^{2}\hbar\omega_{0}^{2}\mathcal{N}_{xyz}}{9\mu^{*2}(4\omega_{0}^{2}-\omega^{2})},\nonumber\\
\chi&=&\frac{-e^{3}}{\mu\mu^{*}(\omega_{0}^{2}-\omega^{2})^{2}},\quad\zeta=\frac{e^{3}\hbar\omega_{0}(4\omega_{0}^{2}-3\omega^{2})\mathcal{N}_{xyz}}
{18\mu^{*3}\omega(4\omega_{0}^{2}-\omega^{2})^{2}},\nonumber\\
\beta&=&\frac{2e^{2}\hbar C\omega^{3}_{0}(\omega^{4}+7\omega_{0}^{2}\omega^{2}+4\omega_{0}^{4})\mathcal{M}_{xyz}}
{\mu^{2}\mu^{*}(\omega^{4}-5\omega_{0}^{2}\omega^{2}+4\omega_{0}^{4})^{3}},\label{labeta}\\
\gamma&=&\frac{-e^{3}\hbar C\omega^{3}_{0}\omega^{2}(\omega^{2}+12\omega_{0}^{2})\mathcal{M}_{xyz}}
{\mu^{2}\mu^{*2}(\omega^{4}-5\omega_{0}^{2}\omega^{2}+4\omega_{0}^{4})^{3}},\\
\xi&=&\frac{2e^{3}\hbar C\mathcal{M}_{xyz}\omega^{3}_{0}\omega(19\omega^{6}-842\omega^{4}\omega_{0}^{2}-224\omega^{2}\omega_{0}^{4}
-672\omega_{0}^{6})}
{15\mu^{2}\mu^{*2}(\omega^{2}-\omega_{0}^{2})^{3}(\omega^{2}-4\omega_{0}^{2})^{5}},\nonumber\\
\varsigma&=&\frac{-4e^{3}\hbar C\mathcal{M}_{xyz}\omega^{3}_{0}\omega(43\omega^{6}-1664\omega^{4}\omega_{0}^{2}-848\omega^{2}\omega_{0}^{4}
-384\omega_{0}^{6})}
{15\mu^{2}\mu^{*2}(\omega^{2}-\omega_{0}^{2})^{3}(\omega^{2}-4\omega_{0}^{2})^{5}},\nonumber
\end{eqnarray}
where $\mathcal{M}_{xyz}$ and $\mathcal{N}_{xyz}$ are dimensionless functions of the anisotropy factors, 
$\mathcal{M}_{xyz}\equiv\eta^{zy}\eta^{yx}\eta^{xz}$, $\mathcal{N}_{xyz}\equiv\eta^{zy}\eta^{yx}+\eta^{xz}\eta^{zy}+\eta^{yx}\eta^{xz}$.
 Using Maxwell's equations 
 the constitutive relations can be written as functions of the electric field alone. In particular, the electric dipole moment reads
\begin{align}
\langle\mathbf{d}^{\omega}\rangle_{\textrm{rot}}&=\alpha_{E}\mathbf{E}_{\omega} -i\omega\chi\mathbf{B}_{0}\wedge\mathbf{E}_{\omega} 
+i\beta\mathbf{k}\wedge\mathbf{E}_{\omega}\\&+(\xi/2-\gamma/\omega)(\mathbf{B}_{0}\cdot\mathbf{k})\mathbf{E}_{\omega}\nonumber\\&+
(\xi/2+\gamma/\omega)(\mathbf{B}_{0}\cdot\mathbf{E}_{\omega})\mathbf{k}.\nonumber
\end{align}
The magnetochiral birefringence comes from the combination of parameters  $\xi/2-\gamma/\omega$ multiplying $\mathbf{B}_{0}\cdot\mathbf{k}$.\\
\indent For a dilute medium, to first order in the number of molecules per unit volume, $\rho$, the macroscopic multipole moment densities are given by,
$\mathbf{P}_{\omega}=\rho\langle\mathbf{d}_{\omega}\rangle_{\textrm{rot}}$, $\mathbf{M}_{\omega}=\rho\langle\mathbf{m}_{\omega}\rangle_{\textrm{rot}}$, 
$\bar{\mathbf{Q}}_{\omega}=\rho\langle\bar{\mathbf{\Theta}}_{\omega}\rangle_{\textrm{rot}}$. Denoting next by $n_{\omega}$ the refractive index of the 
corresponding effective medium, with $n_{\omega}=1+\delta n_{\omega}$, the wave equation for a complex-valued optical field of frequency $\omega$ is 
$n_{\omega}^{2}\mathbf{E}_{\omega}-\epsilon_{0}^{-1}\mathbf{D}_{\omega}=\mathbf{0}$,
with $\mathbf{D}_{\omega}=\epsilon_{0}\mathbf{E}_{\omega}+\mathbf{P}_{\omega}-\textrm{\textonehalf}\:\mathbf{k}\cdot\bar{\mathbf{Q}}_{\omega}-
\omega^{-1}\mathbf{k}\wedge\mathbf{M}_{\omega}$ \cite{Jackson}, 
 from which the values of $n_{\omega}$ can be determined \cite{Ruchon2006}.\\
\indent Our interest is in the relation between the rotatory power, the isotropic refractive index and the magnetochiral birefringence. 
At $\mathcal{O}(\rho)$, the variation in the isotropic refractive index is $\delta n_{0}=\rho\alpha_{E}/2\epsilon_{0}$. The
polarization dependent refractive index due to NOA reads, 
\begin{equation}
\delta n_{NOA}^{\pm}=\pm \frac{\rho\omega}{\epsilon_{0}c}\beta,\label{nbeta}
\end{equation}
and the rotatory power expressed in rad/m is,
\begin{equation}
\varphi=\frac{\pi}{\lambda}(\delta n_{NOA}^{+}-\delta n^{-}_{NOA})=\frac{4\pi^{2}}{\lambda^{2}\epsilon_{0}}\rho\beta,
\end{equation}
where $\delta n^{\pm}_{NOA}$ 
is the variation of the refractive index for left/right circularly polarizaed light. The variation of the refractive index due to magnetochiral birefringence 
depends only on the relative direction between $\mathbf{B}_{0}$ and $\mathbf{k}$ and is independent of the polarization state,
\begin{equation}
\delta n_{MCh}=\frac{\rho}{\epsilon_{0}}(\mathbf{B}_{0}\cdot\mathbf{k})(\xi/2-\gamma/\omega).\label{ngamma}
\end{equation}
From Eq.(\ref{ngamma}), the magnetochiral index reads 
\begin{equation}
\Delta n_{MCh}=\frac{4\pi\rho}{\lambda\epsilon_{0}}\textrm{B}_{0}(\xi/2-\gamma/\omega).\label{la15}
\end{equation}
\indent Far from resonances, in the quasi-static limit $\omega\rightarrow0$ we find,
\begin{align}
\alpha_{E}(0)&=\frac{e^{2}}{\mu\omega_{0}^{2}},\quad\beta(0)=\frac{e^{2}\hbar C\mathcal{M}_{xyz}}{8\mu^{2}\mu^{*}\omega_{0}^{5}},\label{beta0}\\
\quad\xi/2-&\gamma/\omega=\frac{23e^{3}\hbar C\mathcal{M}_{xyz}\omega}{160\mu^{2}\mu^{*2}\omega_{0}^{7}},\quad\omega\rightarrow0.\label{gamma0}
\end{align}
In Eq.(\ref{gamma0}) we note that for $\omega\rightarrow0$ the magnetic term $\gamma/\omega$ is $30/7$ times greater than
the quadrupole term $\xi/2$. We will not attempt to compute the unkown parameters on these quantities from the microscopical data of the spatial configuration 
of charges within specific molecules \cite{Condon2}. Instead, we consider those parameters as effective and we derive their values   
 from experimental macroscopical data. The transition frequency $\omega_{0}$ can be read from $\alpha_{E}(0)$ and so from the isotropic refractive index. The product of 
the anisotropy factors and the strength of the chiral potential, $C\mathcal{M}_{xyz}$, can be read from $\beta(0)$ 
and so from the rotatory power. By comparing the above expressions we find the relation,
\begin{equation}
\quad\xi/2-\gamma/\omega\simeq	\frac{23\omega}{20e}\alpha_{E}(0)\beta(0),\quad\omega\rightarrow0.
\end{equation}
Inserting this relation into Eq.(\ref{la15}) and writing $\alpha_{E}(0)$ and $\beta(0)$ in terms of $\delta n_{0}$ and $\varphi$ at leading order in $\rho$,
$\Delta n_{MCh}$ can also be written as,
\begin{equation}\label{laeq}
\Delta n_{MCh}\simeq\frac{23c\epsilon_{0}}{5e\rho}\textrm{B}_{0}\varphi\delta n_{0},\quad \textrm{at }\mathcal{O}(\rho)\:\textrm{ and }\:\omega\rightarrow0,
\end{equation}
where $\rho$ and $\delta n_{0}$ correspond to the values of the pure compound.\\
\indent We test this formula with the experimental data reported in Ref.\cite{Ruchon2006} for three chiral compounds. The data are cast in Table \ref{table:nonlin}. 
\begin{table}[ht]
\caption{Magnetochiral index of three chiral compounds as predicted from Eq.(\ref{laeq}). In parenthesis, the experimental value from Ref.\cite{Ruchon2006}. 
Weight density, $D$, and refractive index, $n_{0}$, extracted from Refs.\cite{Handbook,Hampton}.} 
\centering 
\begin{tabular}{c c c c c} 
\hline\\
Compound &[$\varphi$]\tiny{[$\frac{\textrm{deg/dm}}{\textrm{g/cm}^{3}}$]}&$n_{0}$ &$D$\tiny{(g/cm$^{3}$)}&$\Delta n_{MCh}[\frac{10^{-10}\textrm{T}^{-1}}{\textrm{g/cm}^{3}}]$
\\ [1.5ex]
\hline\\ 
R-limonene&  &  & & \\[-0.5ex] 
C$_{10}$H$_{16}$& 154 & 1.47 & 0.843 & 2.6 (-3.2)\\\\
L-tartaric acid&  & & &  \\[-0.5ex]
C$_{4}$H$_{6}$O$_{6}$& 12.4 & 1.50 & 1.76 & 0.12 (-0.9) \\\\
L-proline&  & & &  \\[-0.5ex]
C$_{5}$H$_{9}$NO$_{2}$(H$_{2}$O)& -84 &1.3527& - & -0.4 (2.7)\\[-0.5ex]
 &  &\tiny{(1M)}& & \\[1ex] 
\hline 
\end{tabular}
\label{table:nonlin} 
\end{table}
The values predicted by our model are generally smaller than the experimental ones. While the agreement is good for the case of R-limonene, our prediction is
seven times less than the experimental value for the case of L-tartaric. For 
L-proline in aqueous solution the value of $\delta n_{0}/\rho$ was estimated from the optical data of the commercial solution of concentration 1M in Ref.\cite{Hampton}. The 
predicted value for $\Delta n_{MCh}$ is approximately of the same order of magnitude as the experimental value reported in Refs.\cite{Ruchon2006,ValletPRL2001} 
and far from the measurement in Ref.\cite{Weigner1999}, $\sim10^{-8}$T$^{-1}/$ (g/cm$^{3}$).
Nonetheless we notice that higher order terms in $\rho$ and local field factors have been neglected in our calculations.
They may be specially relevant for the case of L-tartaric for which $\rho\sim10^{28}$molecules/m$^{3}$.\\
\indent Next we compare our expression in Eq.(\ref{laeq}) with the formula of Baranova-Zel'dovich (BZ) \cite{Baranova79}. 
The BZ formula, based on the Becquerel's formulation for the Faraday effect, relates the magnetochiral index to the 
frequency dispersion of the refractive index due to NOA, 
\begin{equation}
\Delta n_{MCh}=\frac{2\pi e\textrm{B}_{0}}{\lambda\mu^{*}}\frac{\partial(\delta n_{NOA}^{+}/k)}{\partial\omega}.\label{BZ}
\end{equation}
In this equation the spectral form of $\delta n_{NOA}$ must be calculated either analytically or numerically or estimated from experimental data otherwise. 
Note that our approach resembles that of BZ, with the difference 
that in our case $\Delta n_{MCh}$ depends on $\beta(0)$ and not on its frequency dispersion.\\
\indent If we consider for $\delta n_{NOA}$ in Eq.(\ref{nbeta}) the molecular rotatory factor $\beta(\omega)$ of Eq.(\ref{labeta}) calculated with our model,
we obtain applying Eq.(\ref{BZ}) at leading order in $\rho$, $\Delta n_{MCh}=\frac{4\pi\rho}{\lambda\epsilon_{0}}\textrm{B}_{0}\frac{11e^{3}\hbar C\mathcal{M}_{xyz}\omega}{16\mu^{2}\mu^{*2}\omega_{0}^{7}}$.
This result is approximately five times larger than the value computed in Eq.(\ref{la15}) with $\xi/2-\gamma/\omega$ given by Eq.(\ref{gamma0}).\\
\indent On the contrary, independently of the underlying microscopical model, one can adjust $\varphi$ and thus $\delta n_{NOA}=\lambda\varphi/2\pi$ 
to a fitting function in the form $\varphi= F/(\lambda^{2}-\lambda_{r}^{2})$, where $F$ is a constant and $\lambda_{r}$ is a resonant
wavelength for the chiroptical properties of the molecule \cite{Ruchon2005,Charney}, which is not necessarily the same as the resonance wavelength in the refractive index.
For the case of R-limonene, $\lambda_{r}=180$nm in the function of $\varphi$ \cite{Ruchon2005} and one obtains applying 
Eq.(\ref{BZ}), $\Delta n_{MCh}\simeq3\cdot10^{-11}\textrm{T}^{-1}/(\textrm{g/cm}^{3})$, which is an order of magnitude less than both our predicted and 
the experimental values.\\
\indent Finally we note that both our model and the BZ formula predict equal signs for $\Delta n_{MCh}$ and $\varphi$, while the reported 
experiments in Ref.\cite{Ruchon2006} give opposite signs. This remains to be explained.

\section{Conclusions}
We have carried out a fully analytical quantum calculation of the magnetochiral birefringence based on the single-oscillator model of Condon \emph{et al.} \cite{Condon1,Condon2}.
Taking as given the experimental data of the refractive index and the rotatory power of a dilute molecular medium and demanding consistency with our model, we have 
derived an expression for the magnetochiral refractive index as a function of its isotropic refractive index and its rotatory power [Eq.(\ref{laeq})]. 
In comparison to the approach of Baranova-Zel'dovich, ours is not based on the Becquerel model and predicts a different dispersion behaviour. In fact, the BZ
formula applied to our model yields a factor five in excess for $\Delta n_{MCh}$.\\
\indent The result in Eq.(\ref{laeq}) is in reasonable agreement 
with the experimental values reported. Deviations from the experimental values may be due to the neglect in that equation of higher order terms in $\rho$. 
More fundamentally, it may be due to the oversimplification of the chiral potential $V_{C}$. Additional contributions to NOA may come from
other chromophoric groups within the molecule and from the interaction between their respective magnetic and electric dipole moments \cite{Kirkwood}.\\
\indent Nonetheless, the  
model renders the correct orders of magnitude for $\Delta n_{MCh}$ and hence proves useful for the analytical study of other related physics effects. 
In particular, a similar expression to Eq.(\ref{laeq}) 
was found in Ref.\cite{DonairevTgRikken} using the Condon model for the Casimir kinetic momentum transferred to a chiral molecule 
in a magnetic field, $\mathbf{P}^{\textrm{Cas}}=g\mathbf{B}_{0}$. In this expression $g$ is a pseudo-scalar which presents the same symmetry features as $\xi/2-\gamma/\omega$. 
Therefore, the result obtained here reinforces the conjecture made there \cite{DonairevTgRikken} that the relation of $g$ with $\beta(0)$ and $\alpha_{E}(0)$ is universal up to prefactors 
of order unity.





\bigskip

This work was supported by the ANR contract PHOTONIMPULS
ANR-09-BLAN-0088-01.

\end{document}